\documentclass[a4paper]{spie}  

\usepackage{amsmath,amsfonts,amssymb}
\usepackage{siunitx}
\usepackage{graphicx}
\usepackage[colorlinks=true, allcolors=blue]{hyperref}
\newcommand{\GP}{GRAVITY$^+$}

\title{GRAVITY faint: reducing noise sources in GRAVITY$^+$ with a fast metrology attenuation system}

\author[1]{Felix Widmann}
\author[1]{Stefan Gillessen}
\author[1]{Thomas Ott}
\author[1]{Taro Shimizu}
\author[1]{Frank Eisenhauer}
\author[1]{Maximilian Fabricius}
\author[8]{Julien Woillez}
\author[8]{Frederic Gont\'e}
\author[4]{Matthew Horrobin}
\author[1]{Jingyi Shangguan}
\author[1]{Senol Yazici}
\author[2]{Guy Perrin}
\author[2]{Thibaut Paumard}
\author[3]{Wolfgang Brandner}
\author[3]{Laura Kreidberg}
\author[4]{Christian Straubmeier}
\author[5]{Karine Perraut}
\author[5]{Jean-Baptiste Le Bouquin}
\author[6,7]{Paulo Garcia}
\author[9]{Sebastian H\"onig}
\author[10]{Denis Defr\`ere}
\author[1]{Guillaume Bourdarot}
\author[1]{Antonia Drescher}
\author[1]{Helmut Feuchtgruber}
\author[1]{Reinhard Genzel}
\author[1]{Michael Hartl}
\author[1]{Dieter Lutz}
\author[1]{Nikhil More}
\author[1]{Christian Rau}
\author[1]{Sinem Uysal}
\author[1]{Ekkehard Wieprecht}
\author{GRAVITY+ Collaboration}

\affil[1]{Max-Planck-Institut f\"ur Extraterrestrische Physik (MPE), Gie\ss enbachstra\ss e, Garching bei M\"unchen, Germany}
\affil[2]{LESIA, Observatoire de Paris, PSL Research University, CNRS, Sorbonne Universit\'es, UPMC Univ. Paris 06, Univ. Paris Diderot, Sorbonne Paris Cit\'e, 92195 Meudon Cedex, France}
\affil[3]{Max-Planck-Institut f\"ur Astronomie, K\"onigstuhl 17, 69117 Heidelberg, Germany}
\affil[4]{1. Physikalisches Institut, Universit\"at zu K\"oln, Z\"ulpicher Str. 77, 50937 K\"oln, Germany}
\affil[5]{Univ. Grenoble Alpes, CNRS, IPAG, 38000 Grenoble, France}
\affil[6]{CENTRA and Universidade de Lisboa - Faculdade de Ciencias, Campo Grande, 1749-016 Lisboa, Portugal}
\affil[7]{Faculdade de Engenharia, Universidade do Porto, rua Dr. Roberto
Frias, 4200-465 Porto, Portugal}
\affil[8]{European Southern Observatory, Karl-Schwarzschild-Str. 2, 85748 Garching, Germany}
\affil[9]{School of Physics \& Astronomy, University of Southampton,
Southampton, SO17 1BJ, UK}
\affil[10]{Institute of Astronomy, KU Leuven, Celestijnenlaan 200D, B-3001,
Leuven, Belgium}

\authorinfo{Further author information:\\F.Widmann.: E-mail: fwidmann@mpe.mpg.de}

\begin{document} 
\maketitle

\begin{abstract}
With the upgrade from GRAVITY to GRAVITY+ the instrument will evolve to an all-sky interferometer that can observe faint targets, such as high redshift AGN. Observing the faintest targets requires reducing the noise sources in GRAVITY as much as possible. The dominant noise source, especially in the blue part of the spectrum, is the backscattering of the metrology laser light onto the detector. To reduce this noise we introduce two new metrology modes. With a combination of small hardware changes and software adaptations, we can dim the metrology laser during the observation without losing the phase referencing. For single beam targets, we can even turn off the metrology laser for the maximum SNR on the detector. These changes lead to a SNR improvement of over a factor of two averaged over the whole spectrum and up to a factor of eight in the part of the spectrum currently dominated by laser noise.
\end{abstract}


\section{INTRODUCTION}
\label{sec:intro}  
Since the first light of GRAVITY in 2016 \cite{Gravity2017}, GRAVITY and the VLTI have transformed optical interferometry with groundbreaking results in many different science areas. With the recent success comes an increased demand for optical interferometry and the wish to push the limits of what is observable even further. This is exactly the goal of the project \GP{}. \GP{} aims to improve GRAVITY and the existing VLTI infrastructure to transform GRAVITY into an all-sky interferometer with increased sensitivity \cite{Eisenhauer2019}. The upgrades involved in \GP{} are implemented in a phased approach and several improvements have already been completed. This includes a replacement of the grisms inside GRAVITY to improve the throughput \cite{Yazici2021}, as well as the implementation of wide field fringe tracking, i.e. GRAVITY-Wide \cite{gravity2022}. Other parts of the project, such as a completely new AO system with laser guide stars for all VLTI UTs and improved vibration control, are currently under development. With the goal of creating a more sensitive GRAVITY instrument comes also the task to eliminate existing noise sources as much as possible. One of these noise sources is the backscattering of the metrology laser, which completely dominates the noise in the blue part of the science spectrum. The goal of the project presented here, which is called GRAVITY-faint, is to develop an observing mode in which we can minimize the noise from the laser without losing the functionality of the metrology system.

\section{The GRAVITY metrology system}
\begin{figure}
\centering
\includegraphics[width=0.7\textwidth]{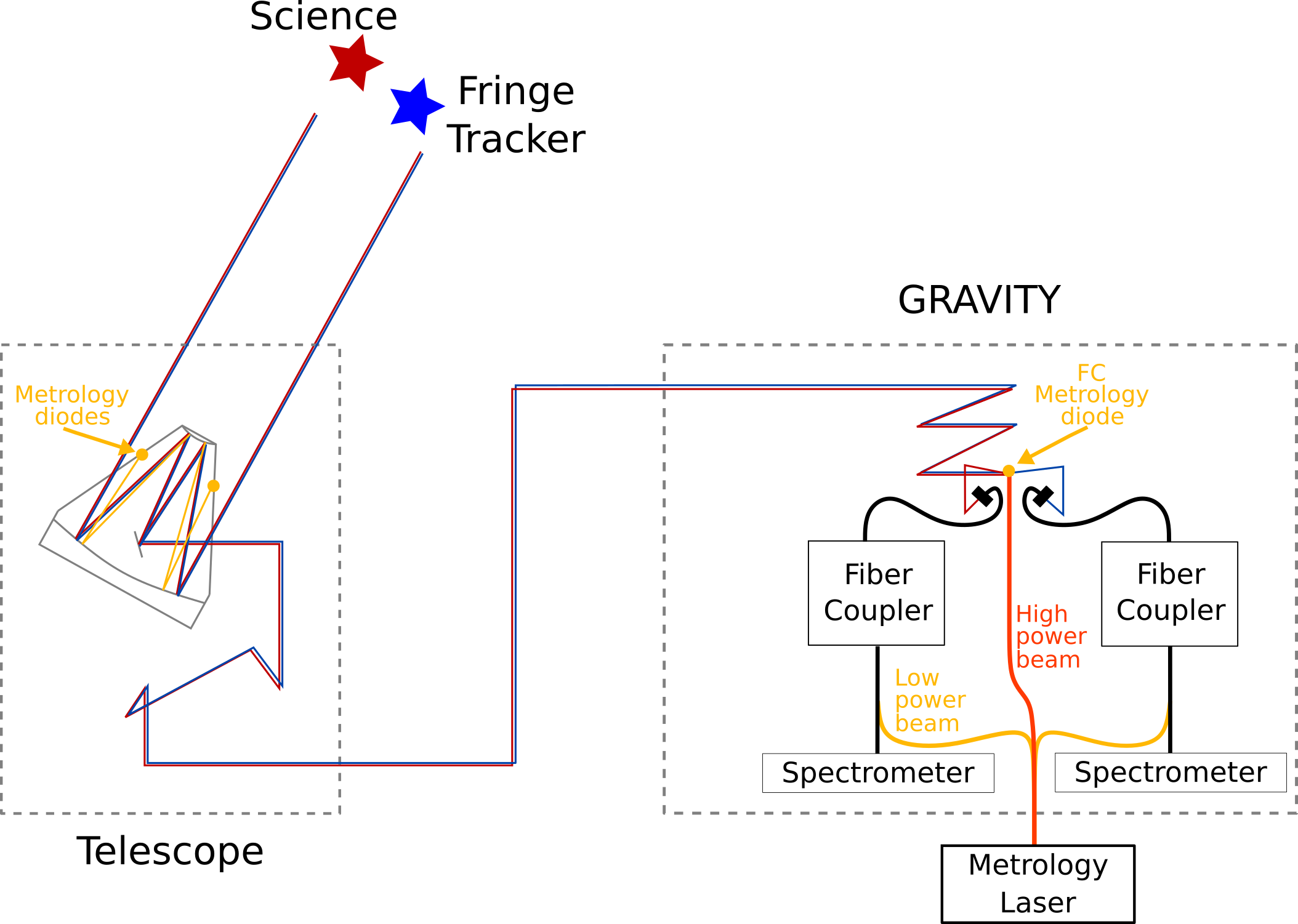}
\caption{Schematic overview of the metrology system used by GRAVITY. On the right, GRAVITY is shown with the metrology beam split up into the two low power beam which are fed into the two fiber couplers and the high power beam which is fed into the system after the fiber couplers. The signal is then measured in the telescope (shown on the left) with four metrology diodes, which are mounted on the telescope spiders. For simplicity only one telescope is shown, and many optical elements (such as the delay lines) are not shown.}
\label{fig:metrology} 
\end{figure} 

GRAVITY offers the possibility to measure precise distances between two objects in the dual-field mode \cite{Gravity2017}. In this mode one object is the fringe tracking (FT) star, and one is the science (SC) target. The astrometry is done by measuring the differential optical path difference (dOPD) between the two targets. To measure this dOPD, one has to ensure that the internal dOPD between the two beams in the interferometer is well known. Measuring the instrumental dOPD is the task of the GRAVITY metrology system \cite{Lippa2014}. The metrology system consists of three beams from the same laser, which are propagated backwards through the light-path of GRAVITY and the VLTI. By splitting up one laser one ensures fixed phase-relations between the three beams. Two beams with low laser power are fed into both of the GRAVITY beam combiners (one for the FT and one for the SC beam). The third beam, which has a significantly higher laser power, is fed into the system after the fiber coupler. This beam is referred to as the carrier beam. For a schematic overview of the system, see \autoref{fig:metrology}.

GRAVITY uses this three-beam scheme to avoid feeding the high power beam through any fiber optics and, therefore, to minimize the noise of the metrology laser in the science spectrum. While the laser's wavelength, \SI{1908}{\nano\meter}, is outside the wavelength range of the science spectrum (\SIrange{2000}{2500}{\nano\meter}), Raman scattering and fluorescence were found to occur in the optical fibers, which scatter the laser light into the science wavelengths. This backscattering leads to significant noise in the science spectrum and would completely dominate over any science signal if the high power laser would be directly fed into the beam combiners. The backscattering happens on rare elements in the optical fibers and cannot be avoided. For a more detailed description of the problem see Ref. \citenum{Lippa2018}. With the three-beam scheme the backscattering is reduced, as only the low power beams are fed into the fibers. However, it is not completely removed and is still a major noise source for GRAVITY.

After the fiber coupler all three beams follow the starlight backwards and are measured in the telescope with four metrology diodes. The diodes are mounted on the spiders of each telescope and measure individually the dOPD between the carrier beam and each of the FT and SC beams. The measurement is done with photodiodes and the three beams are disentangled by phase-shifting each of them with a characteristic frequency. The dOPD is measured by comparing the FT-carrier and the SC-carrier measurement. The carrier is not needed for the final measurement but is used to amplify the signal from the two faint beams.

With this system, the metrology traces all the optical paths from GRAVITY back to the primary mirror of the telescopes. The measurement above the M1 is performed in the pupil plane, which sets the narrow angle baseline for astrometric measurements \cite{Lacour2014}.

Another function of the metrology is to control motion inside the GRAVITY instrument. For this, there is another metrology diode in the beam combiner, which measures the same signal as the telescope diodes. However, this so-called fiber coupler (FC) diode has a much higher SNR \cite{Pfuhl2014}. The dOPD measurement from the FC diode is used to control the movement of the fibered differential delay lines (FDDLs) in GRAVITY, which corrects for the dOPD between the two targets in dual-field observing mode introduced by the sky motion. As the FDDLs are operated in closed loop with the FC metrology signal, this has to be considered when changing to different metrology modes.

\section{Noise analysis}
\begin{figure}
\centering
\includegraphics[width=0.6\textwidth]{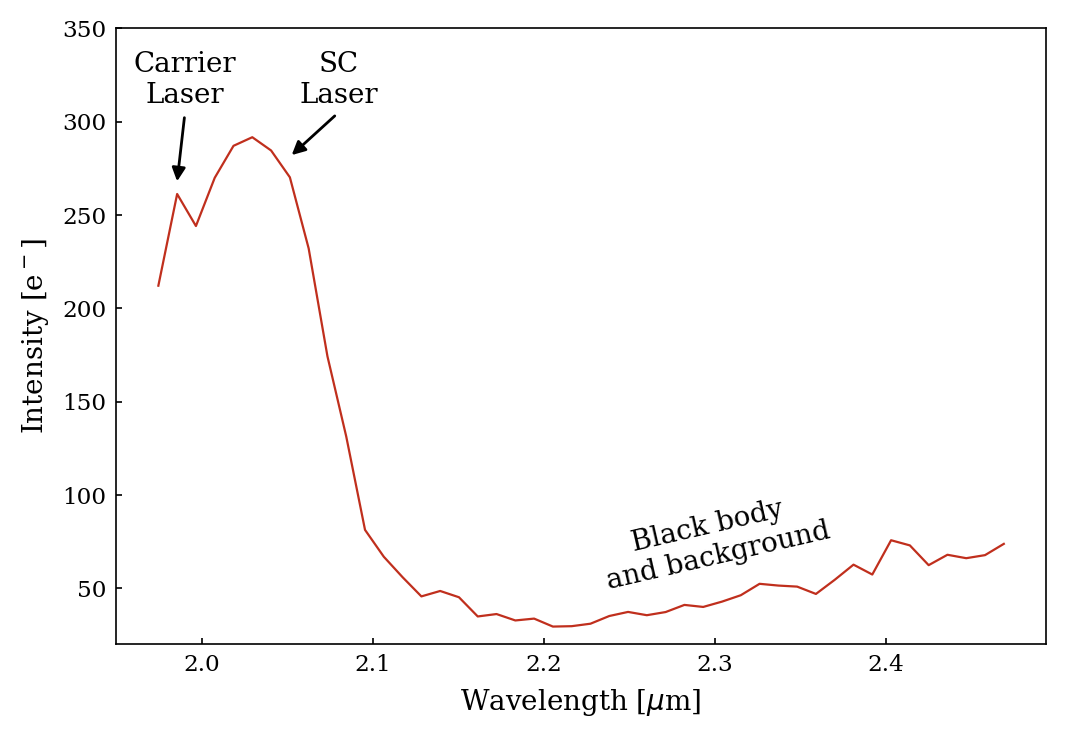}
\caption{Measured intensity on a \SI{10}{\second} dark frame in medium resolution with an indication for the dominant sources over the spectrum.}
\label{fig:intensity} 
\end{figure} 
The noise on the SC detector of GRAVITY has several different contributions. The normal detector readout noise and the dark current are comparably low. The GRAVITY science detector is a HAWAII2RG with a double correlated read noise of approximately 11 electrons. This is consistent with the expected behaviour of the detector \cite{Finger2008}. The noise sources which dominate over the detector noise are from photon noise. This has mainly two components: firstly the thermal background from the telescope and VLTI optical train, and secondly the backscattering of the metrology laser \cite{Widmann2018}. The laser itself has again several components. There is the main broad backscattering which occurs mostly from \SIrange{2.0}{2.1}{\micro\meter}. This is caused by the SC and FT metrology beam in the optical fiber. Additionally we have backscattering of the carrier beam, which causes a peak at wavelengths below \SI{2.0}{\micro\meter} as well as a diffuse background over the full detector, originating from the metrology injection in the spectrometer. The diffuse background has a signal of about one electron per pixel and second, while the backscattering below \SI{2.1}{\micro\meter} is on the order of tens of electrons per pixel and second, depending on the observing mode used. For an example of the measured intensity on a typical dark exposure see \autoref{fig:intensity}. The metrology laser is therefore clearly the dominating noise source with an approximate intensity of 30 electrons per second and a noise of up to six electrons per second. Observations of faint targets would benefit greatly from removing this noise source.

\section{The new metrology modes}
\begin{figure}
\centering
\includegraphics[width=0.7\textwidth]{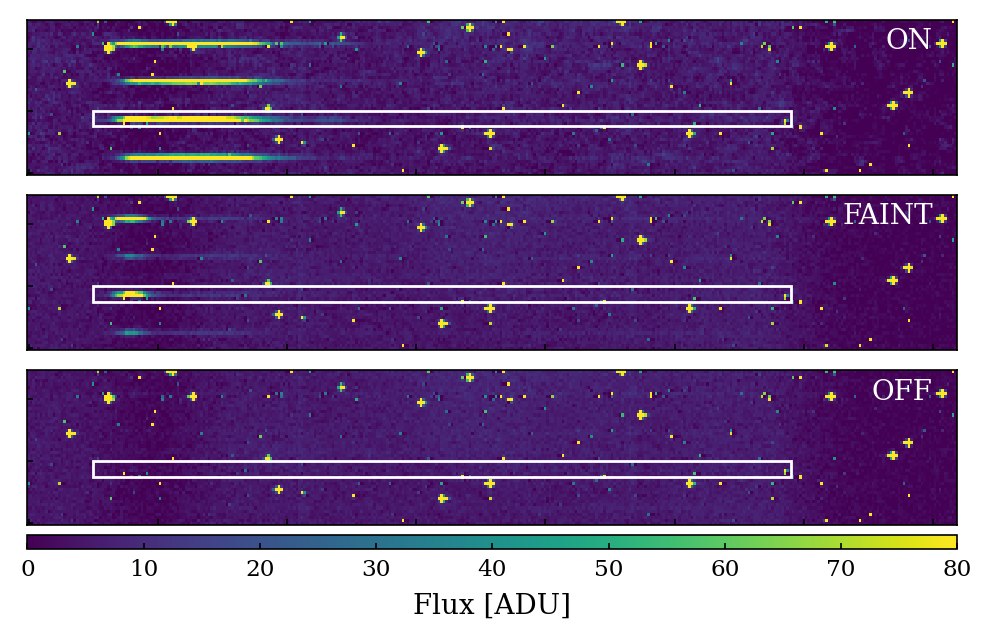}
\caption{Cut out from the detector image of a dark frame in all three modes. Top: metrology on, where clear backscattering can be seen in left part of the spectra. Middle: metrology faint, with only a small peak from the carrier laser remaining. Bottom: metrology off, no spectra visible anymore. The white rectangle in all plots shows a typical extraction region for a spectrum.}
\label{fig:met_det} 
\end{figure} 
To remove metrology laser noise we made a small hardware adjustment to the metrology system of GRAVITY. After the splitting of the metrology laser we added optical attenuators to the SC and FT beam. The attenuators allow us to damp the laser power in the SC and FT beam by applying a voltage to them. The carrier laser stays untouched. However, we implemented a new possibility to shut off the laser amplifier, which effectively shuts off the metrology laser completely. 
With these new additions we can now use the laser metrology system in three different modes:
\begin{itemize}
    \item \textbf{ON:} The laser metrology is used as before
    \item \textbf{FAINT:} The SC and FT are damped during a science exposure to reduce noise on the detector
    \item \textbf{OFF:} The laser is shut off completely to fully remove the metrology noise
\end{itemize}
To illustrate the impact \autoref{fig:met_det} shows the detector image for a dark frame in the three modes. On the GRAVITY SC detector the wavelength axis is on the x-axis. The spectra from the different baselines and polarizations are projected with a shift in y-direction leading to 48 spectra visible as horizontal lines. In \autoref{fig:met_det} we show a cut-out of the detector which includes four spectra. The top plot with metrology ON shows significant backscattering, which then reduces for GRAVITY FAINT. In FAINT the only metrology light left is the very sharp peak at the bluest part of the spectrum, which is from the carrier shining back into the fiber coupler (see for comparison \autoref{fig:intensity}). In the bottom plot in \autoref{fig:met_det} this is then completely gone, as also the carrier is turned off in the metrology OFF mode.

Depending on the science goals of the observation and the expected brightness of the target a different mode should be used. The two new modes are described in the following.

\subsection{Metrology FAINT}
\begin{figure}
\centering
\includegraphics[width=0.95\textwidth]{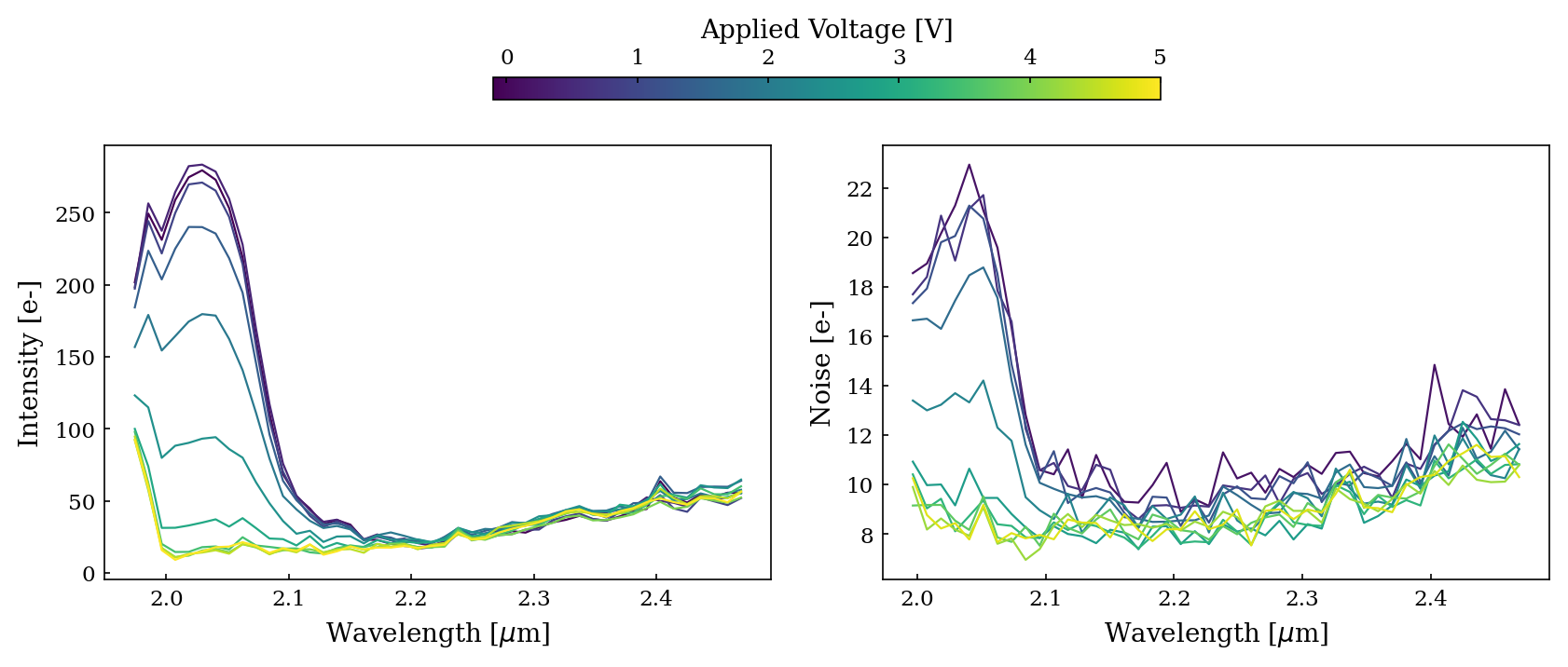}
\caption{Measured intensity on a \SI{10}{\second} dark frame for different voltages applied to the optical attenuators. The voltage ranges from \SI{0}{\volt} (no damping) to \SI{5}{\volt} (fully damped). The left plot shows the measured intensity and the right plot the noise over 10 DITs, estimated by the standard deviation.}
\label{fig:faint_volt} 
\end{figure} 
In the FAINT mode, the main idea is to have the metrology measurement with ideal SNR only in between exposures. This is done by using the attenuators to dampen the metrology signal during the exposure. The remaining intensity on a typical dark frame with different damper voltages is shown in \autoref{fig:faint_volt}. The backscattering, which is the dominant noise source, is removed by damping the light. The scattered light from the laser injection is also reduced with increasing voltage. Due to the bias removal in the reduction of the dark frames this is hardly visible in the measured intensity. However, the right plot in \autoref{fig:faint_volt} shows a clear noise improvement over the full spectrometer. For the fully dampened laser only the sharp peak below \SI{2}{\micro\meter}, which comes from reflections of the carrier laser, as well as the thermal background, are visible. 

\begin{figure}
\centering
\includegraphics[width=0.8\textwidth]{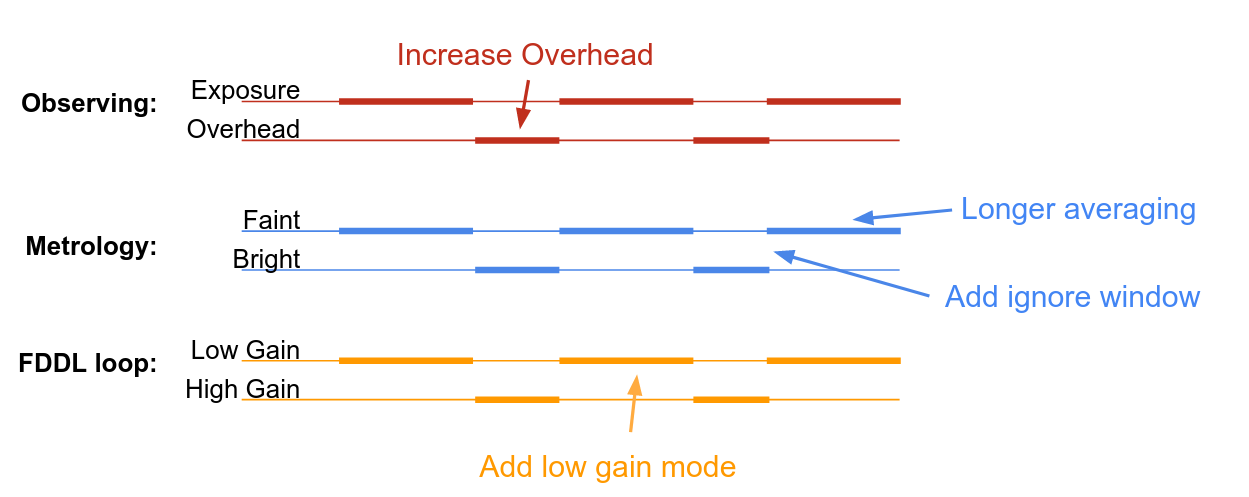}
\caption{Schematic of the FAINT observing mode, which shows the necessary changes that are introduced for this mode.}
\label{fig:faint_mode} 
\end{figure} 

To keep the basic functionality of the metrology system, we damp down the metrology signal to only about \SI{90}{\percent}, which corresponds to a voltage of \SI{3.2}{\volt}. This allows us to always have a continuous metrology signal by using a longer averaging of the measurement. Additionally, we need to use a lower gain for the control loop of the FDDLs, as it relies on the metrology measurement. In this setup with the dimmed metrology, the longer averaging, and the lower gain on the FDDLs, we keep the stability of the instrument with a significantly improved SNR on the detector. Additionally, there is a wait time of \SI{1}{\second} introduced in between exposures. In this \SI{1}{\second} the metrology laser is brought to the original brightness by removing the voltage from the attenuators and the normal averaging and gain factors are used. This allows the FDDLs to move to their nominal position in case they would show small deviations from it. The last change is, that we introduce a short ignore time around the changes of the attenuator, where the metrology signal is simply ignored. This avoids a wrong measurement in the case that the change in voltage at the attenuators introduces a phase error. The changes and the observing principle is schematically shown in \autoref{fig:faint_mode}.

\begin{figure}
\centering
\includegraphics[width=0.8\textwidth]{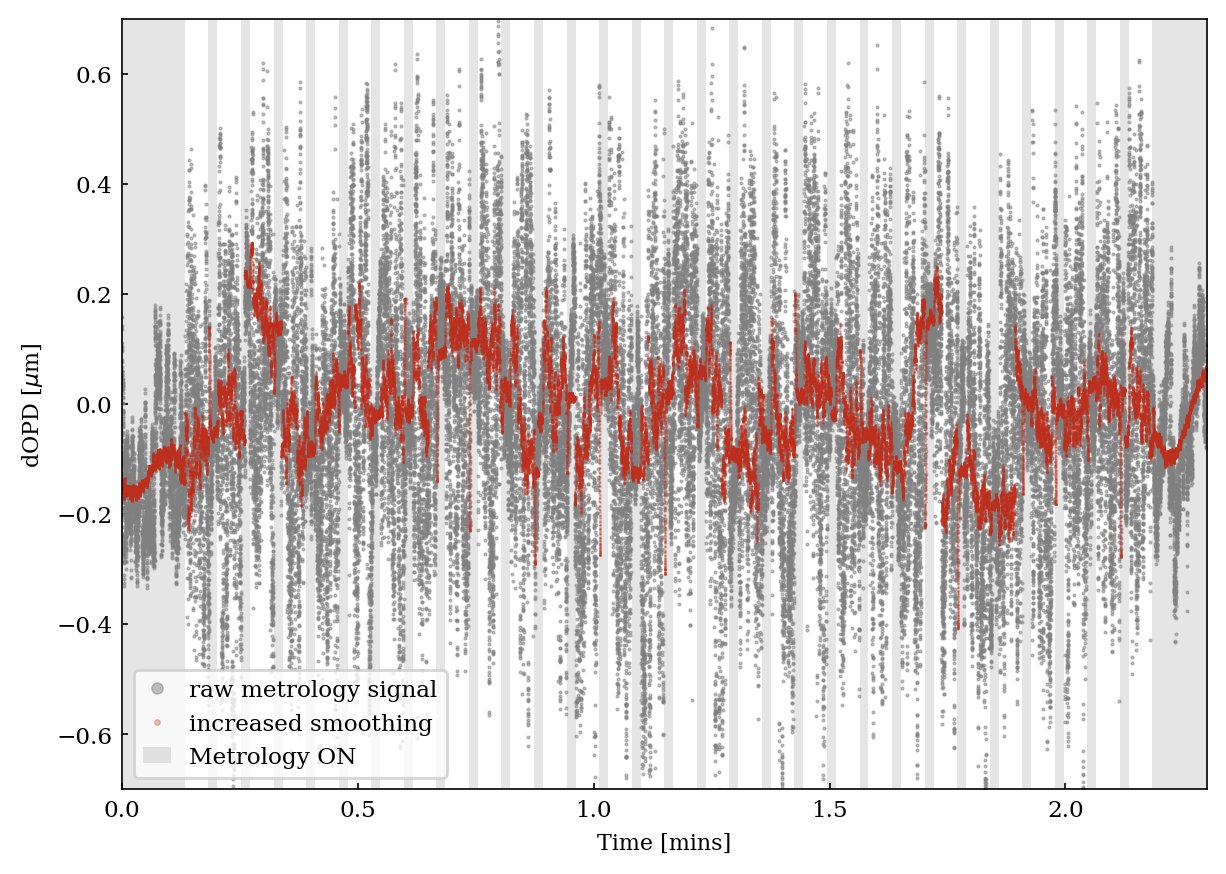}
\caption{Metrology signal in GRAVITY-faint. The grey dots show the metrology measurement as it is used in normal mode. The red data show the same measurement with increased smoothing. The grey areas in the background indicate times in which the metrology was bright.}
\label{fig:gfaint_met} 
\end{figure}
\begin{figure}
\centering
\includegraphics[width=0.98\textwidth]{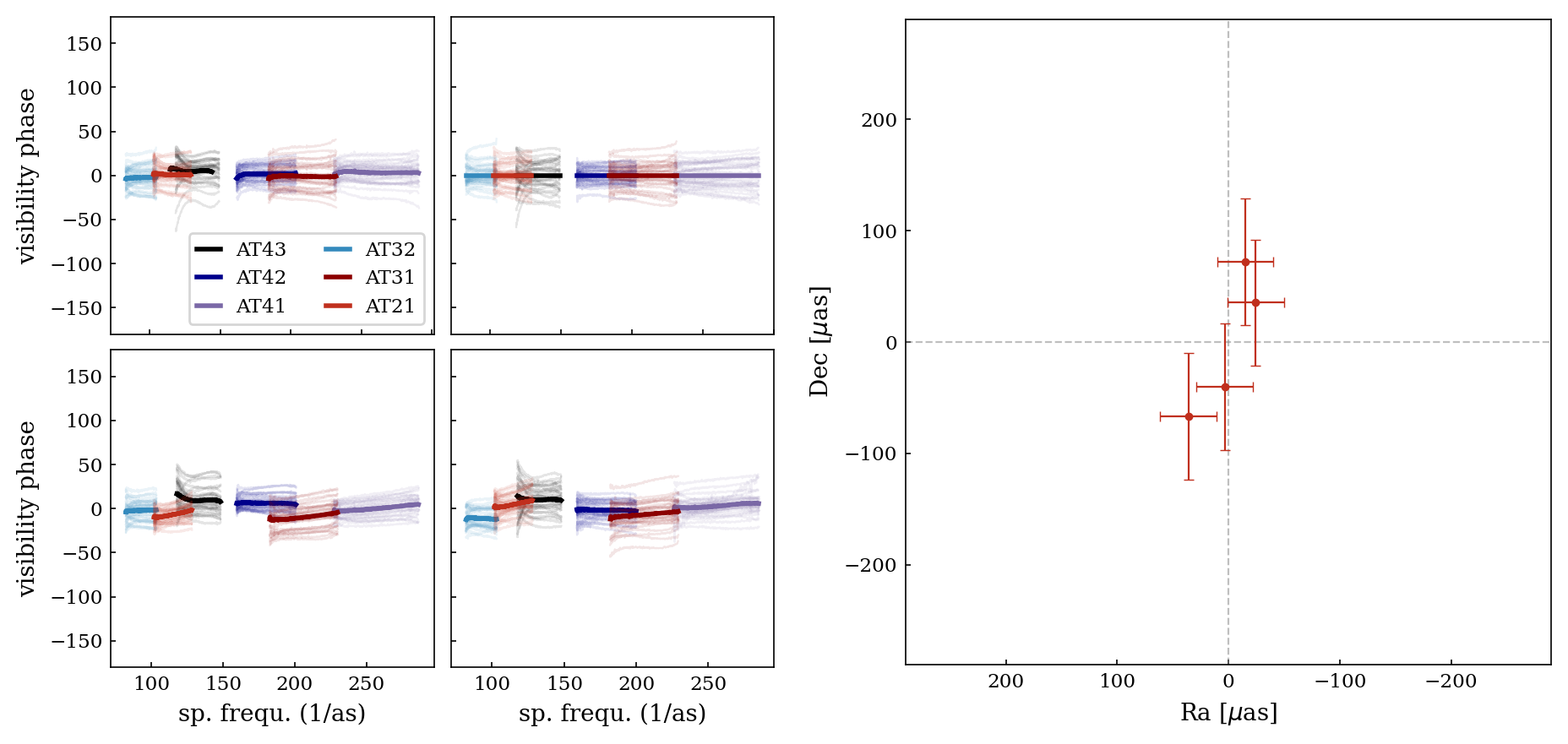}
\caption{First test of an astrometric measurement with GRAVITY-faint. The left plots show the visibility phase over 4 exposures of \SI{2}{\min} each, with the average phase as a bold line and the phase per DIT shaded out. The right plot shows the astrometry of the four files, relative to the calibrated phase center.}
\label{fig:gfaint_pos} 
\end{figure} 
Similar to the real-time code we also introduce a longer averaging in the data reduction software, which then results in the metrology signal at a nominal SNR. The metrology signal for one exposure and one telescope is shown in \autoref{fig:gfaint_met}, where we show that we can get the high SNR metrology measurement back by introducing a longer averaging.

To show the stability of the metrology system in faint mode, we observed a binary target during the commissioning. This target was observed in normal dual-field mode, where one star is the FT target and the other one the SC. The resulting SC visibility phases are shown in \autoref{fig:gfaint_pos}. The phases are calibrated with the expected phase center of the SC target, so they should be zero for a single source. The four exposures in \autoref{fig:gfaint_pos} do indeed show phases of around zero, the small scatter comes from noise and aberration in the instrument. The right plot in \autoref{fig:gfaint_pos} shows the fitted position of the SC star, which is again roughly consistent with zero and shows an expected scatter.
This is so far only a very first test, as we need to further optimize the new parameters of the system and run longer tests to assess the stability, but the first results look promising.

\subsection{Metrology OFF}
\begin{figure}
\centering
\includegraphics[width=0.9\textwidth]{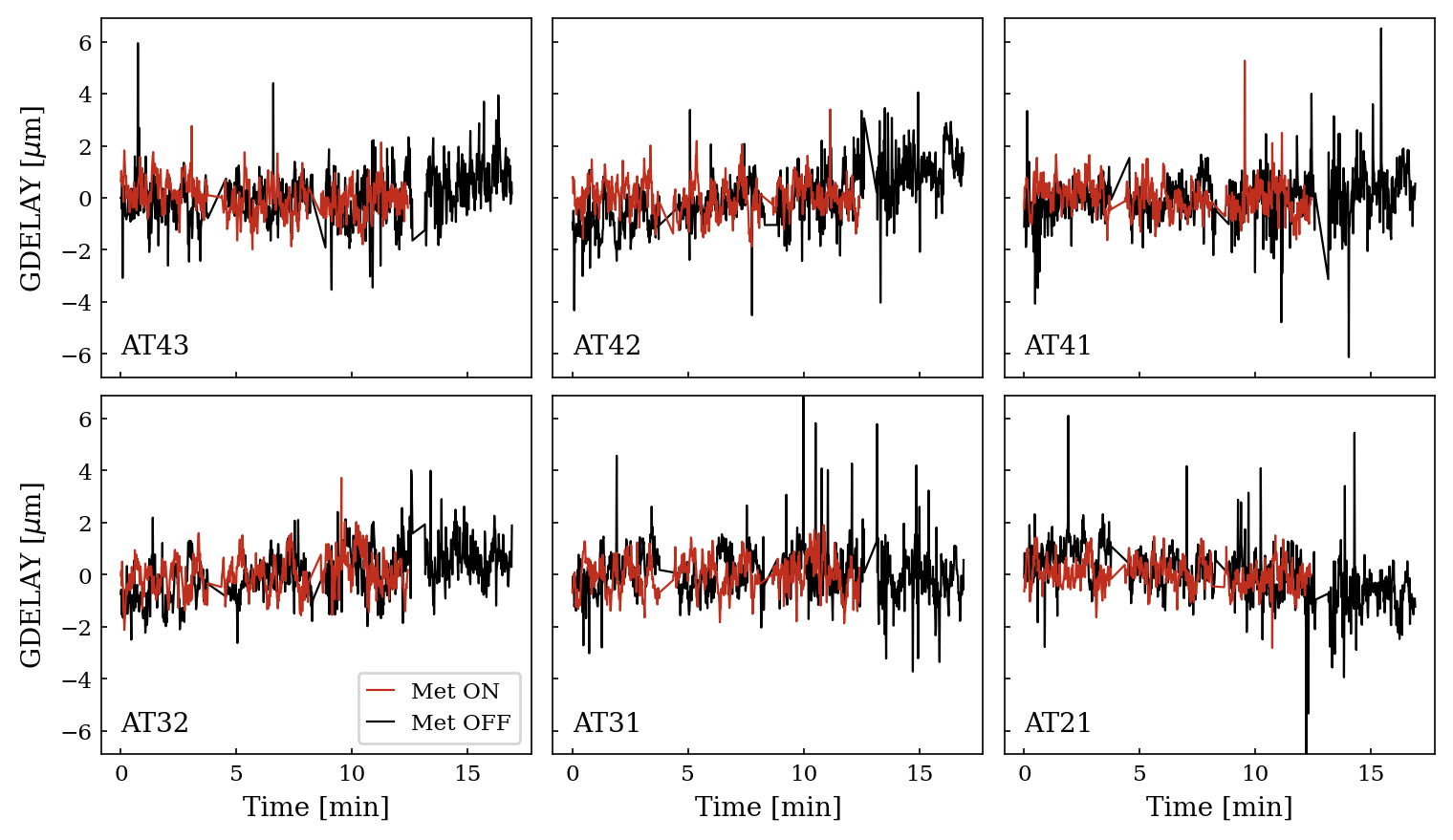}
\caption{Group delay measurement for a star pair with metrology on (red) and metrology off (black).}
\label{fig:gdelay} 
\end{figure} 
The metrology measurement is crucial to allow dual-field astrometry with GRAVITY. However, many science cases of GRAVITY are based on single beam measurements and therefore do not depend on metrology measurements. So far the metrology noise was not a big issue in single beam cases, as for those observations usually, the FT star is the same as the SC star, which sets a limit of roughly 10 mag to observations with the UTs. For such bright targets, the photon noise from the target dominates over all other noise sources.

With the implementation of GRAVITY-Wide \cite{gravity2022} as part of \GP{} the number of targets with the possibility of off-axis fringe tracking significantly increases, which opens up observations of many faint science targets. For these targets, it is therefore also interesting to suppress the metrology noise.

For faint targets, we introduce a second new mode which is the metrology OFF mode. In this mode, the metrology laser is completely turned off by removing the power from the laser amplifier. This happens during the acquisition and the laser stays permanently turned off. This means that not only the SC and FT laser are shut off, but also the carrier beam. Therefore, the full laser noise including the sharp carrier peak as well as the diffuse backscattering over the full detector vanishes.

For targets that do not depend on dual-field astrometry, the main purpose of the metrology signal is to keep the instrument in a stable state and avoid drifts in the OPD. From tests in the commissioning of GRAVITY-Wide, we could demonstrate that the instrument is indeed stable without the metrology feedback. For these tests, we observed a star pair in GRAVITY-Wide in both metrology ON and OFF. In \autoref{fig:gdelay} the group delay of the two measurements is shown and there is no visible drift in the group delay of the metrology OFF measurement, which shows that the instrument is indeed stable. The data in \autoref{fig:gdelay} spans only over \SI{15}{\min}. We took much more data in metrology OFF mode, for example, the observation of a QSO at redshift z=2  in Ref. \citenum{gravity2022} was taken with metrology OFF and showed a stable instrument over timescales of hours.

\subsection{Remaining noise}
\begin{figure}
\centering
\includegraphics[width=0.6\textwidth]{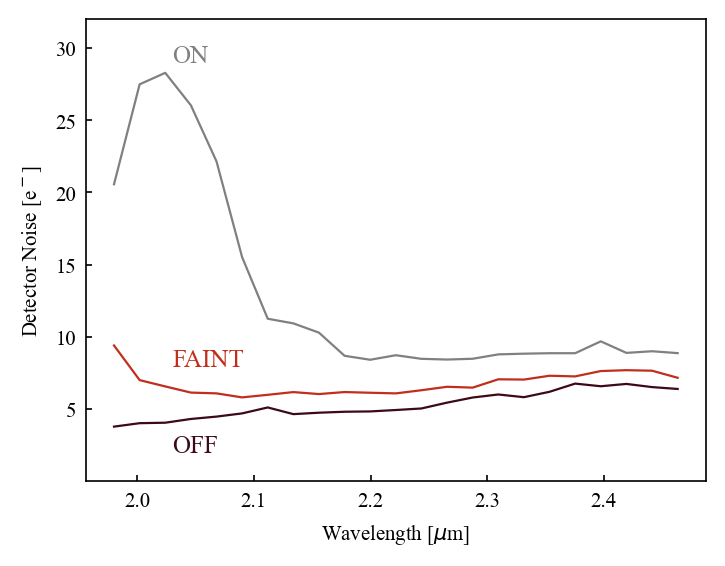}
\caption{Comparison of the detector noise in the three different metrology mode. The noise is measured on dark frames with \SI{30}{\second} integration.}
\label{fig:all_noise}
\end{figure}
With the two new modes, we could significantly improve the noise introduced by the metrology system in GRAVITY. An overview of the remaining noise profile in a typical dark frame is shown in \autoref{fig:all_noise}. We have the clearest improvement in metrology OFF mode, where the dominating noise terms are now the detector noise as well as the thermal emission from the telescope and VLTI optical train. The noise decreases by a factor of eight in the blue part of the spectrum (from \SIrange{2.0}{2.1}{\micro\meter}) and on average by a factor of 2.5. The improvement in the FAINT mode is slightly lower, as the metrology carrier beam which is responsible for the noise at the lowest wavelengths stays untouched in this mode. Nevertheless, also the FAINT mode shows significant improvement while still allowing for phase-referenced dual-field observations.

\section{Conclusion}
In this paper, we present our work to suppress the noise from the metrology laser in GRAVITY, which is part of the \GP{} project. We can clearly demonstrate the improvements with the two new modes, which will ultimately allow getting the best possible SNR for faint targets with \GP{}. With the FAINT mode, we achieve this while still keeping the metrology signal and allowing for fully phase-referenced dual-field observations. This mode is coming close to its completion and will be offered for observations with GRAVITY in the future. The second mode is the metrology OFF mode which will give the best possible SNR for science cases that do not rely on dual-field metrology measurements. This mode is already available now and was successfully used for the first science demonstrations of the GRAVITY-Wide mode \cite{gravity2022}.


\bibliography{report}
\bibliographystyle{spiebib}

\end{document}